\documentclass[aps, prd, amsmath, floats, floatfix, twocolumn, nofootinbib]{revtex4-1}
\usepackage{graphicx}
\usepackage{color}
\usepackage{latexsym}

\newcommand{\be}{\begin{eqnarray}}
\newcommand{\ee}{\end{eqnarray}}

\begin{document}

\title{Shadows of CPR black holes and tests of the Kerr metric}

\author{M. Ghasemi-Nodehi}

\author{Zilong Li}

\author{Cosimo Bambi}
\email[Corresponding author: ]{bambi@fudan.edu.cn}

\affiliation{Center for Field Theory and Particle Physics and Department of Physics, Fudan University, 200433 Shanghai, China}

\date{\today}

\begin{abstract}
We study the shadow of the Cardoso-Pani-Rico (CPR) black hole for different values of the black hole spin $a_*$, the deformation parameters $\epsilon_3^t$ and $\epsilon_3^r$, and the viewing angle $i$. We find that the main impact of the deformation parameter $\epsilon_3^t$ is the change of the size of the shadow, while the deformation parameter $\epsilon_3^r$ affects the shape of its boundary. In general, it is impossible to test the Kerr metric, because the shadow of a Kerr black hole can be reproduced quite well by a black hole with non-vanishing $\epsilon_3^t$ or $\epsilon_3^r$. Deviations from the Kerr geometry could be constrained in the presence of high quality data and in the favorable case of a black hole with high values of $a_*$ and $i$. However, the shadows of some black holes with non-vanishing $\epsilon_3^r$ present peculiar features and the possible detection of these shadows could unambiguously distinguish these objects from the standard Kerr black holes of general relativity.
\end{abstract}

\maketitle


\section{Introduction \label{s-1}}

Astrophysical black hole (BH) candidates are compact objects in X-ray binaries with a mass $M \approx 5-20$~$M_\odot$ and supermassive bodies at the center of galaxies with a mass $M \sim 10^5 - 10^{10}$~$M_\odot$~\cite{nara}. In the framework of standard physics, the spacetime geometry around these objects should be described by the Kerr solution of general relativity. However, an observational confirmation is still lacking and deviations from standard predictions can be motivated by a number of arguments, from classical extensions of general relativity~\cite{v2-1,v2-2,v2-3,v2-4} to macroscopic quantum effects~\cite{gia1,gia2,gidd}.

The electromagnetic radiation emitted by the gas in the inner part of the accretion disk propagates through the strong gravitational field of these objects and it is inevitably affected by relativistic effects that carry information about the spacetime geometry and the nature of BH candidates. The study of the properties of this radiation can potentially test the Kerr BH paradigm~\cite{rev1,rev2}.

Today, the two main techniques to probe the spacetime geometry around BH candidates are the study of the thermal spectrum of thin disks (continuum-fitting method)~\cite{cfm1,cfm2,cfm3} and the analysis of the iron K$\alpha$ line~\cite{iron1,iron2}. These techniques are normally used to measure the spin parameter of BH candidates under the assumption of the Kerr background, but they can be generalized to non-Kerr metrics to constrain possible deviations from the Kerr solution~\cite{harko1,harko1.5,harko2,harko3,k1,k2,k3,k4,k5}. The thermal spectrum of thin disk has a simple shape and therefore it is fundamentally impossible to test the Kerr metric~\cite{kcfm1,kcfm2}. The iron line profile has a more complicated structure and it is thus potentially a more powerful tool, but high quality data would be necessary~\cite{jjc1,jjc2}. Current observations can rule out some black hole alternatives, like some kinds of exotic dark stars~\cite{exotic0,exotic1} and some types of wormholes~\cite{exotic2}, but more motivated scenarios are quite difficult to test.

Since the problem of testing the Kerr metric is related to the strong correlation between the estimate of the spin and possible deviations from the Kerr geometry, it is natural to try to combine several measurements of the same candidate with the goal to break the degeneracy among the parameters of the metric. Unfortunately, this is not possible at the moment, because the continuum-fitting method and the iron K$\alpha$ line are both mainly sensitive to the position of the inner edge of the disk~\cite{cfmiron}, while other measurements are not yet mature to test fundamental physics~\cite{qpo0,qpo1,qpo2,jet1,jet2}.

SgrA$^*$ is the supermassive BH candidate at the center of the Galaxy and an ideal laboratory to test general relativity in the near future. This object is 5-6 orders of magnitude more massive than any other BH candidate in the Galaxy and much closer than the other supermassive BH candidates in galactic nuclei. While there are currently no available observations to test the geometry around SgrA$^*$, new observational facilities may soon provide unprecedented measurements. Sub-millimeter very long baseline interferometry facilities should be able to resolve the BH shadow, namely to measure the apparent photon capture radius of SgrA$^*$~\cite{falke,sh1,sh2,sh3,sh4,sh5,sh6,sh7,sh7.5,sh8,sh9,sh10}. High frequency radio observations may find pulsars in close orbits around SgrA$^*$ and then get an unambiguous measurement of its spin parameter $a_*$~\cite{pulsar}. The spin parameter may also be measured with normal stars~\cite{stars}. High resolution observations of blobs of plasma orbiting SgrA$^*$ may be soon available and provide additional information on the spacetime geometry around this object~\cite{spot1,spot2,spot3}. The spectrum of the accretion structure, if properly understood, may become another important tool to test the Kerr metric~\cite{nan}.

In this paper, we want to further investigate the possibility of testing the Kerr metric with the detection of the BH shadow. In this sense, the natural target for this kind of observations is SgrA$^*$, but in principle the same observations could be possible for any BH candidate surrounded by an optically thin emitting medium. For instance, another good candidate is the supermassive BH in M87, which is more distant than SgrA$^*$ but also more massive, and eventually its angular size in the sky should be only slightly smaller than that of SgrA$^*$. Here we employ the Cardoso-Pani-Rico (CPR) metric~\cite{cpr} and we study the shadow of these BHs in terms of their spin parameter $a_*$, deformation parameters $\epsilon_3^t$ and $\epsilon_3^r$, and viewing angle $i$ (namely the angle between the the spin and the line of sight of the observer). $\epsilon_3^t$ mainly changes the size of the shadow, since it regulates the gravitational strength. $\epsilon_3^r$ determines the horizon and it alters the shape of the shadow, especially on the side of corotating photon orbits. We find that the shadow of a Kerr BH can be usually reproduced quite well by a non-Kerr BH with non-vanishing deformation parameters and different spin, so the sole measurement of the shadow in general cannot test the Kerr metric. In the case of a fast-rotating Kerr BH observed from a large viewing angle $i$, the situation is better, and it may be possible to constrain $\epsilon_3^t$ and $\epsilon_3^r$. We also find that there are non-Kerr BHs with a qualitatively different shadow: the detection of similar shadows could unambiguously rule out the Kerr metric. In the general case, the shadow measurement may constrain an allowed region on the spin parameter -- deformation parameter plane and the Kerr metric can be tested if we add additional independent measurements.

The content of the paper is as follows. In Sec.~\ref{s-2}, we briefly review the concept of BH shadow. In Sec.~\ref{s-3}, we describe our approach: we use the CPR background and we describe the BH shadow in terms of a certain function $R$. In Sec.~\ref{s-4}, we present the original part of this work: we study the shadow of CPR BHs and the possible observational constraints on $\epsilon_3^t$ and $\epsilon_3^r$ in the presence of a detection. We summarize our results in Sec.~\ref{s-5}. Throughout the paper, we employ units in which $G_{\rm N} = c = 1$ and the convention of a metric with signature $(-+++)$.

\section{Black hole shadow \label{s-2}}

When a BH is surrounded by an optically thin emitting medium, the apparent image of the accretion flow close to the compact object presents a dark area over a bright background~\cite{falke}. Such a dark area is the so-called BH shadow and its boundary corresponds to the photon capture sphere as seen by the distant observer. While the intensity map of the image depends on the properties of the accretion structure and the emission mechanisms, the boundary of the shadow is only determined by the spacetime metric and the viewing angle of the observer. An accurate measurement of the BH shadow can thus be used to test the Kerr BH hypothesis and the shadows of several non-Kerr BHs have been already studied~\cite{sh1,sh2,sh3,sh4,sh5,sh6,sh7,sh7.5,sh8,sh9,sh10}.

The apparent photon capture sphere can be calculated in the following way. We consider the image plane of the distant observer with Cartesian coordinates $(X,Y)$. We fire a photon from every point of the image plane. The photon must have the 3-momentum perpendicular to the image plane. The photon trajectory is numerically integrated from the observer to the BH. Some photons approach the BH and then they escape to infinity. Other photons are captured by the BH and cross the event horizon. The boundary of the shadow is the closed curve on the image plane of the distant observer separating scattered photons from captured photons. In reality, photons are emitted by the medium surrounding the BH and reach the detector at infinity, but for the calculations it is more convenient to proceed in the opposite way and start from the image plane of the observer.

If we use a coordinate system $(t,r,\theta,\phi)$ to describe the BH metric, the initial conditions $(t_0, r_0, \theta_0, \phi_0)$ for the photon with Cartesian coordinates $(X,Y)$ on the image plane of the distant observer are given by~\cite{k2}
\be\label{eq-1}
t_0 &=& 0 \, , \nonumber\\
r_0 &=& \sqrt{X^2 + Y^2 + D^2} \, , \nonumber\\
\theta_0 &=& \arccos \frac{Y \sin i + D \cos i}{\sqrt{X^2 + Y^2 + D^2}} \, , \nonumber\\
\phi_0 &=& \arctan \frac{X}{D \sin i - Y \cos i} \, .
\ee
and the initial conditions for the photon 4-momentum are
\be\label{eq-2}
k^r_0 &=& - \frac{D}{\sqrt{X^2 + Y^2 + D^2}} |\bf{k}_0| \, , \nonumber\\
k^\theta_0 &=& \frac{\cos i - D \frac{Y \sin i + D 
\cos i}{X^2 + Y^2 + D^2}}{\sqrt{X^2 + (D \sin i - Y \cos i)^2}} |\bf{k}_0| \, , \nonumber\\
k^\phi_0 &=& \frac{X \sin i}{X^2 + (D \sin i - Y \cos i)^2} |\bf{k}_0| \, , \nonumber\\
k^t_0 &=& \sqrt{\left(k^r_0\right)^2 + r^2_0  \left(k^\theta_0\right)^2
+ r_0^2 \sin^2\theta_0  (k^\phi_0)^2} \, . 
\ee
We note that, without loss of generality, we assume that the $X$-axis is parallel to the axis of symmetry of the shadow and perpendicular to the BH spin. In our calculations, the observer is located at $D = 10^6$~$M$, which is far enough to assume that the background geometry is flat. $k^t_0$ is thus obtained from the condition $g_{\mu\nu}k^\mu k^\nu = 0$ with the metric tensor of a flat spacetime. With the initial conditions~(\ref{eq-1}) and (\ref{eq-2}), one can numerically solve the geodesic equations to check whether the photon fired from the point $(X,Y)$ hits the BH or not. In the case of the Kerr metric, it is not necessary to solve the geodesic equations, because the spacetime is of Petrov type D and the equations of motions in Boyer-Lindquist coordinates are separable and of first order. However, this is not possible in general, and for this reason here we use the general approach valid for any spacetime.

\section{Theoretical framework \label{s-3}}

\subsection{CPR metric}

Tests of the Schwarzschild metric in the weak field limit are commonly and conveniently discussed within the PPN (Parametrized Post-Newtonian) formalism, see e.g. Ref.~\cite{will}. The idea is to write the most general line element for a static and spherically symmetric spacetime in terms of an expansion in $M/r$, namely
\be\label{eq-ppn}
ds^2 &=& - \left(1 - \frac{2 M}{r} + \beta\frac{2 M^2}{r^2} + . . .  \right) dt^2 \nonumber\\
&& + \left(1 - \gamma \frac{2 M}{r} + . . .  \right) \left(dx^2 + dy^2 + dz^2 \right) \, ,
\ee
where $\beta$ and $\gamma$ are two coefficients that parametrize our ignorance. In the isotropic coordinates of Eq.~(\ref{eq-ppn}), the Schwarzschild solution has $\beta = \gamma = 1$. In Solar System experiments, one assumes that $\beta$ and $\gamma$ are free parameters to be determined by observations. Today we know that $\beta$ and $\gamma$ are 1 with an accuracy at the level of $10^{-4} - 10^{-5}$ and this confirms the validity of the Schwarzschild solution in the weak field limit~\cite{will}.

The same approach can be used to test the Kerr metric, even if the picture is now more complicated because the spacetime is only stationary and axially symmetric, and we are not in the weak field limit any more. At the moment there is not a satisfactory formalism to test the geometry around BH candidates and this may generate some confusion. In any case, the idea is the same: we consider a more general solution that includes the Kerr metric as a special case, and possible deviations from the Kerr geometry are quantified by a set of ``deformation parameters''. Astrophysical observations should measure the values of these deformation parameters and check whether they vanish, so that the metric of the spacetime reduces to the usual Kerr metric of general relativity.

In this paper, we employ the CPR metric~\cite{cpr}. In Boyer-Lindquist coordinates, the line elements reads
\begin{widetext}
\be\label{eq-m}
\hspace{-0.8cm}
ds^2 &=& - \left(1 - \frac{2 M r}{\Sigma}\right)\left(1 + h^t\right) dt^2
 - 2 a \sin^2\theta \left[\sqrt{\left(1 + h^t\right)\left(1 + h^r\right)} 
- \left(1 - \frac{2 M r}{\Sigma}\right)\left(1 + h^t\right)\right] dt d\phi \nonumber\\
&& + \frac{\Sigma \left(1 + h^r\right)}{\Delta + h^r a^2 \sin^2\theta} dr^2
+ \Sigma d\theta^2
+ \sin^2\theta \left\{\Sigma + a^2 \sin^2\theta \left[ 2 \sqrt{\left(1 + h^t\right)
\left(1 + h^r\right)} - \left(1 - \frac{2 M r}{\Sigma}\right)
\left(1 + h^t\right)\right]\right\} d\phi^2 \, , 
\ee
\end{widetext}
where $a = a_* M$ is the specific BH spin with the dimension of $M$, $\Sigma = r^2 + a^2 \cos^2 \theta$, $\Delta = r^2 - 2 M r + a^2$, and
\be
h^t &=& \sum_{k=0}^{+\infty} \left(\epsilon_{2k}^t 
+ \epsilon_{2k+1}^t \frac{M r}{\Sigma}\right)\left(\frac{M^2}{\Sigma}
\right)^k\, , \\
h^r &=& \sum_{k=0}^{+\infty} \left(\epsilon_{2k}^r
+ \epsilon_{2k+1}^r \frac{M r}{\Sigma}\right)\left(\frac{M^2}{\Sigma}
\right)^k \, .
\ee
There are two infinite sets of deformation parameters, $\{\epsilon_k^t\}$ and $\{\epsilon_k^r\}$ (at increasingly high order). Since the lowest order deformation parameters are already strongly constrained to recover the Newtonian limit and meet the PPN bounds (see~\cite{cpr} for more details), in what follows we consider the deformation parameters $\epsilon_3^t$ and $\epsilon_3^r$. Higher order deformation parameters are instead neglected for the sake of simplicity, even if these terms could be important for fast-rotating BHs.

\begin{figure}[b]
\begin{center}
\includegraphics[type=pdf,ext=.pdf,read=.pdf,width=5cm]{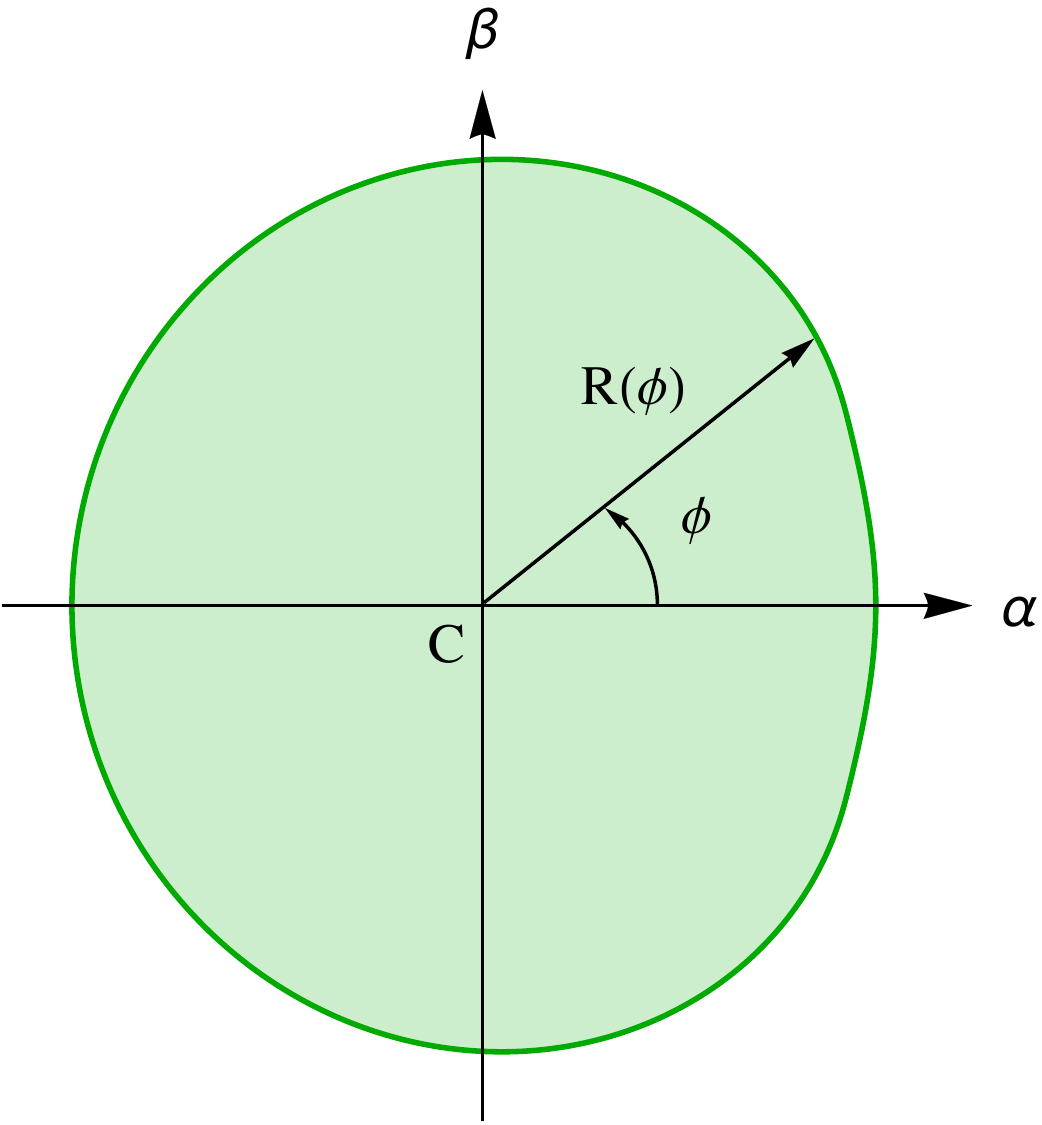}
\end{center}
\caption{The function $R(\phi)$ is defined as the distance between the center C and the boundary of the shadow at the angle $\phi$ as shown in this picture. See the text for more details.}
\label{fig1}
\end{figure}

\subsection{Description of the shadow \label{sub}}

In the next section, we want to compare shadows computed for different values of the parameters of the model, namely $a_*$, $\epsilon^t_3$, $\epsilon^r_3$, and $i$. It is thus necessary a way to describe the shape of the shadow. We employ the method discussed in Ref.~\cite{thindisk} (a similar and more sophisticated approach is proposed in~\cite{rezzolla}).

First, we find the ``center'' C of the shadow. Its Cartesian coordinates on the image plane of the observer are 
\be
X_{\rm C} &=& \frac{\int \rho(X,Y) X dX dY}{\int \rho(X,Y) dX dY} \, , \nonumber\\
Y_{\rm C} &=& \frac{\int \rho(X,Y) Y dX dY}{\int \rho(X,Y) dX dY} \, ,
\ee
where $\rho (X,Y) = 1$ inside the shadow and $\rho (X,Y) = 0$ outside. The shadow is symmetric with respect to the $X$-axis and we can define as $R(0)$ the shorter segment between C and the shadow boundary along the $X$-axis. Defining the angle $\phi$ as shown in Fig.~\ref{fig1}, $R(\phi)$ is the distance between the point C and the boundary at the angle $\phi$. The function $R(\phi)/R(0)$ completely characterizes the shape of the BH shadow. Here we consider $R(\phi)/R(0)$ instead of $R(\phi)$ because the latter cannot be measured with good precision, as it would require an accurate measurement of the distance and the mass of the BH, which are usually not easy. Even the exact positions of the shadow on the image plane of the observer cannot be used to test the Kerr metric, because it is difficult to precisely identify the center $X=Y=0$ of the source.

\begin{figure*}
\begin{center}
\includegraphics[type=pdf,ext=.pdf,read=.pdf,width=8.5cm]{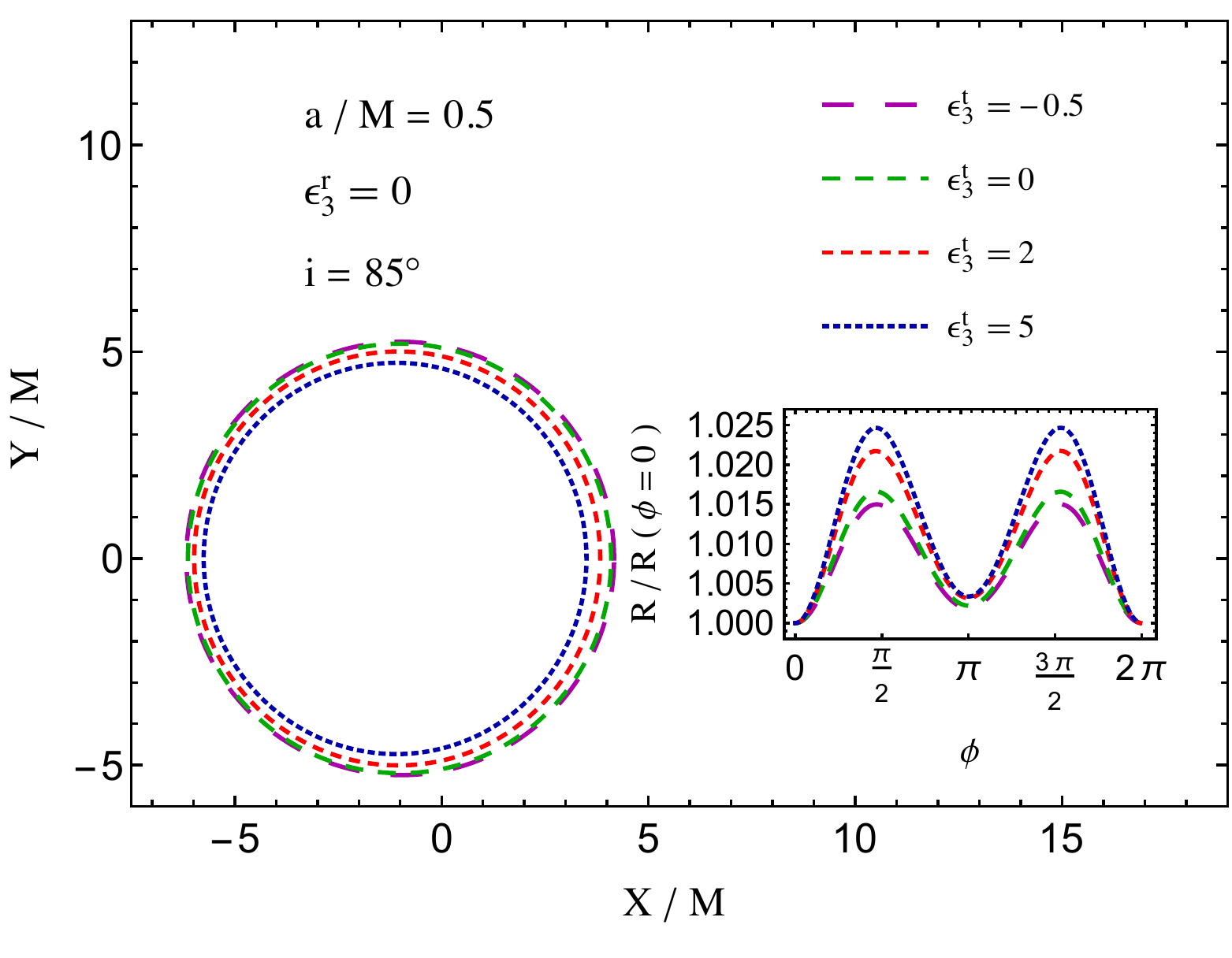}
\hspace{0.5cm}
\includegraphics[type=pdf,ext=.pdf,read=.pdf,width=8.5cm]{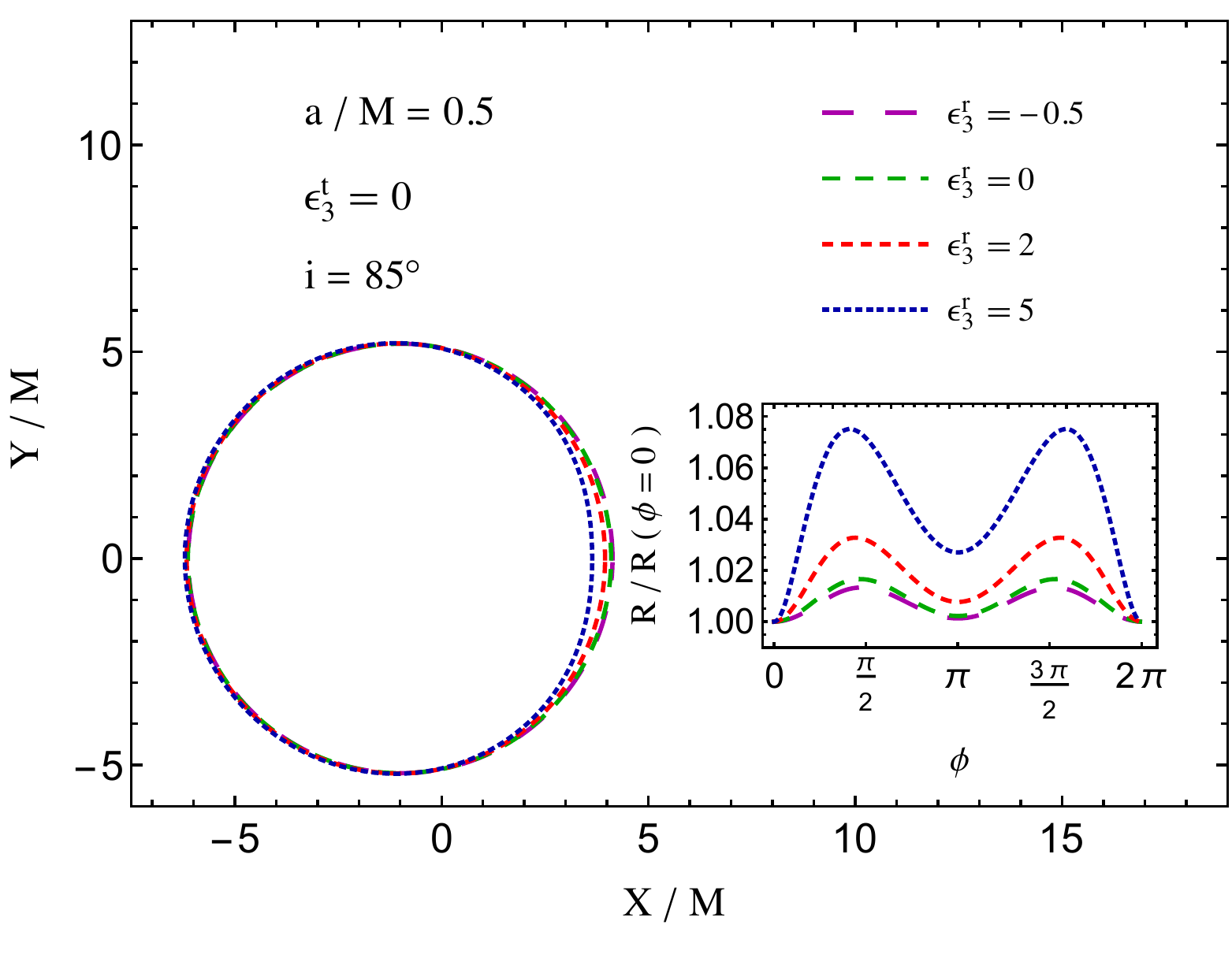} \\
\vspace{0.5cm}
\includegraphics[type=pdf,ext=.pdf,read=.pdf,width=8.5cm]{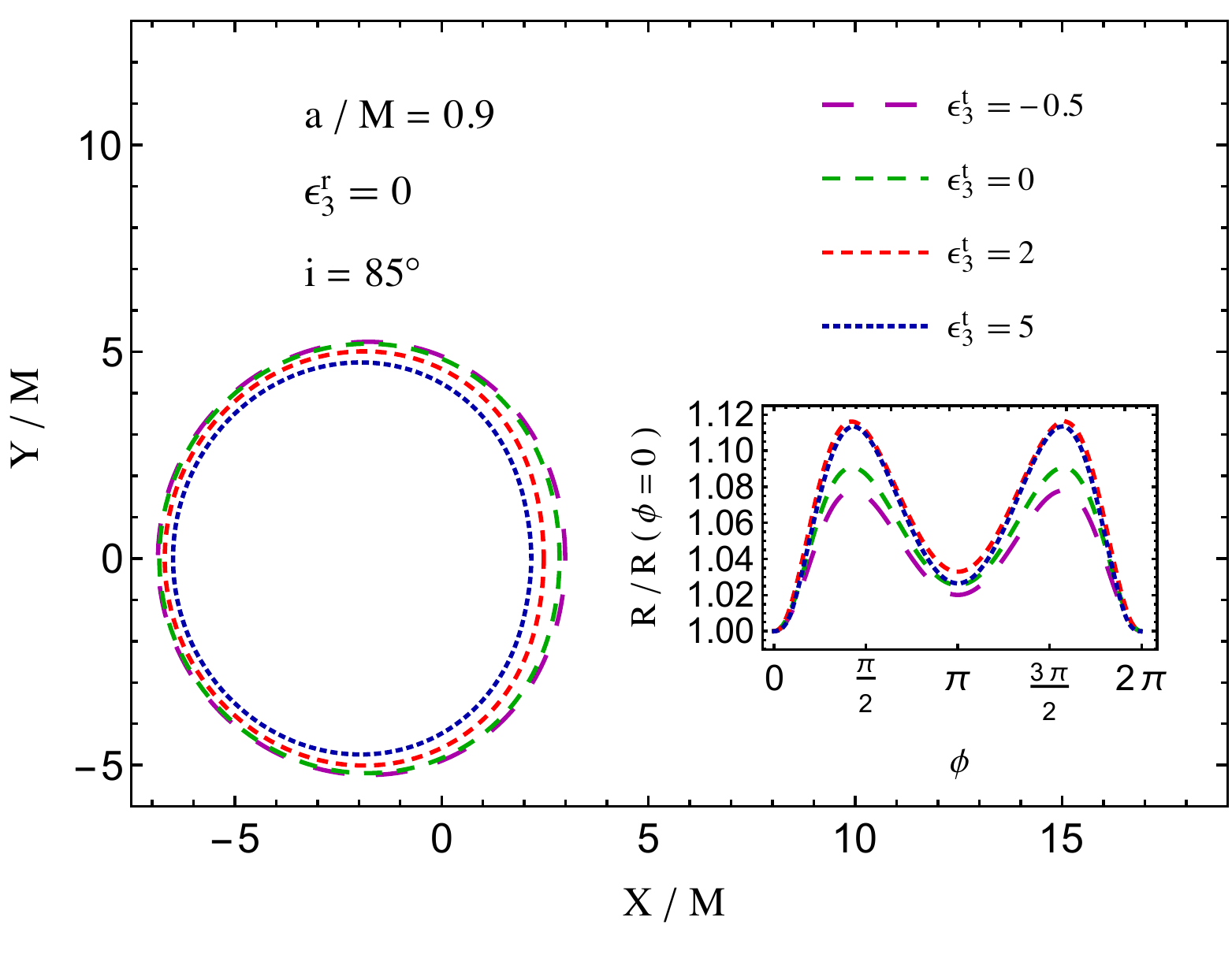}
\hspace{0.5cm}
\includegraphics[type=pdf,ext=.pdf,read=.pdf,width=8.5cm]{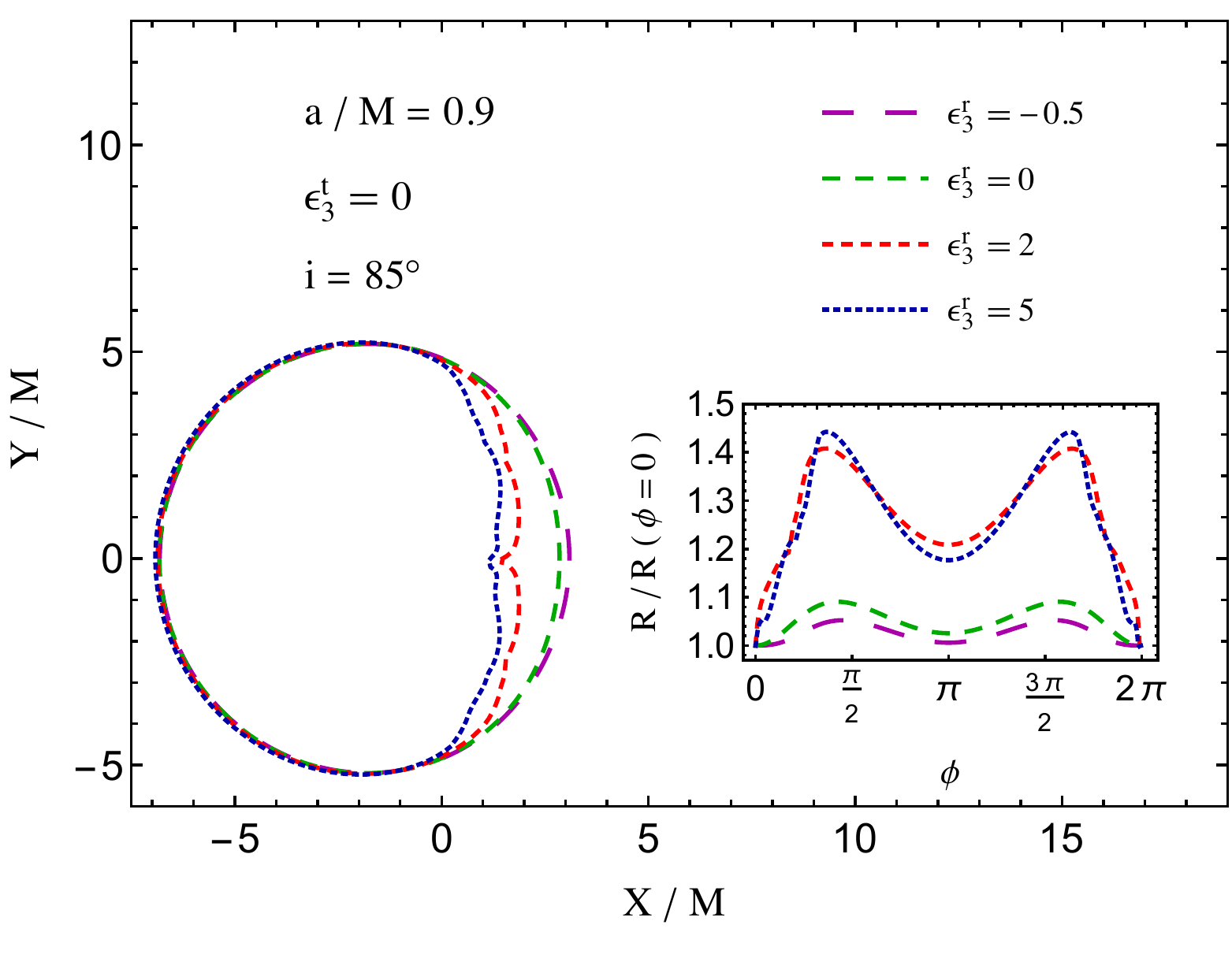} \\
\vspace{0.5cm}
\includegraphics[type=pdf,ext=.pdf,read=.pdf,width=8.5cm]{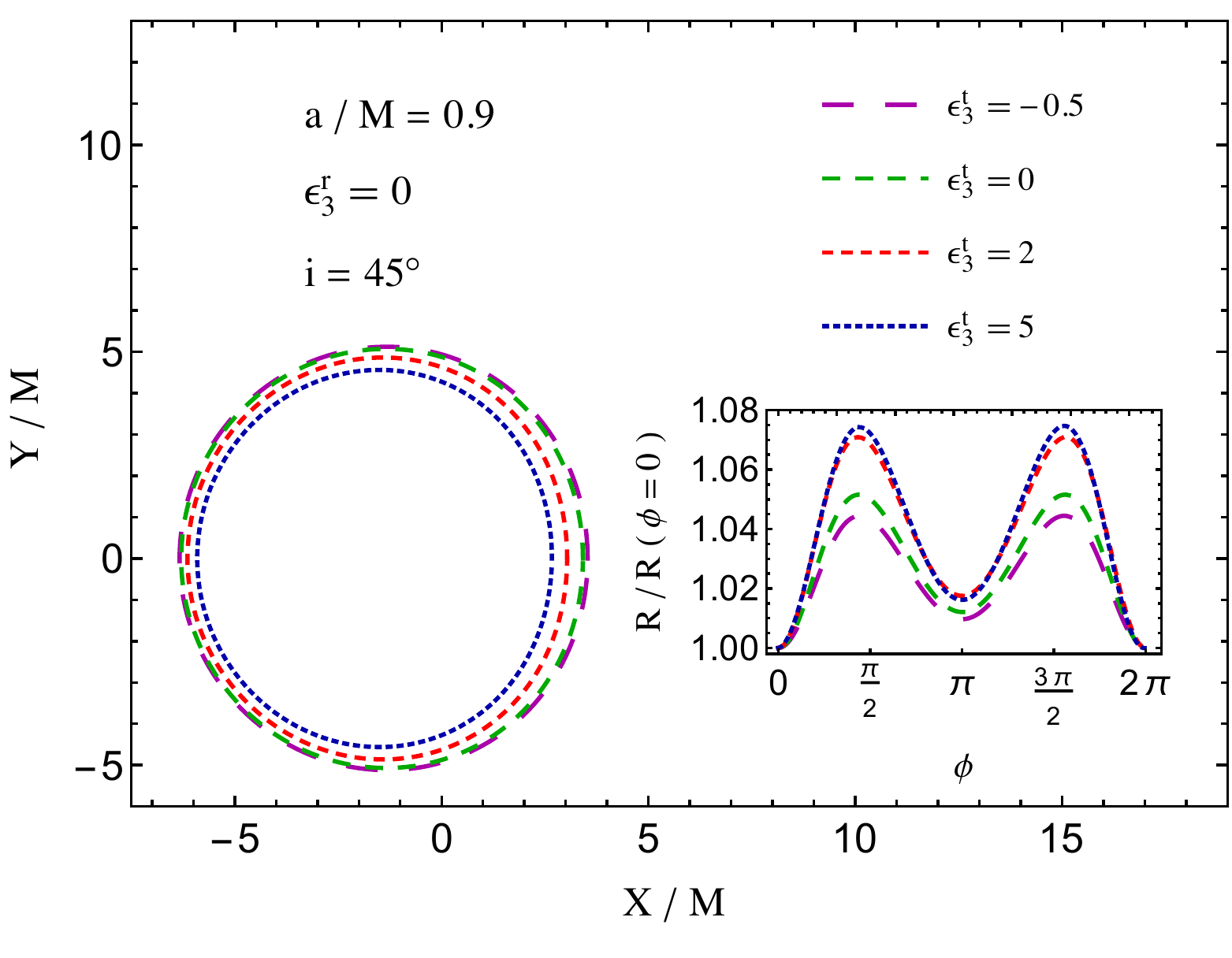}
\hspace{0.5cm}
\includegraphics[type=pdf,ext=.pdf,read=.pdf,width=8.5cm]{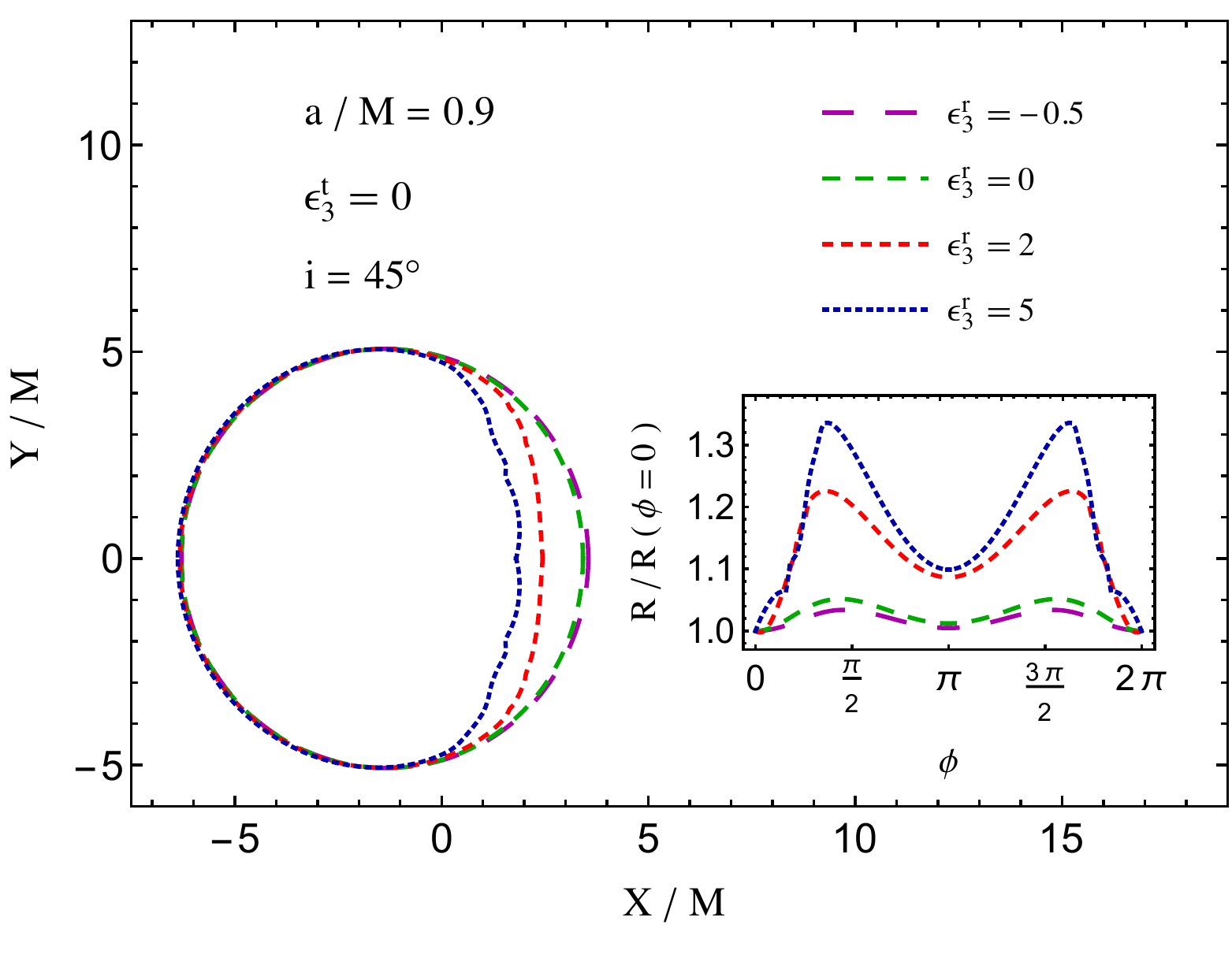} \\
\end{center}
\caption{BH shadows and $R$ functions for CPR BHs with different values of the spin parameter $a_*$, the deformations parameters $\epsilon^t_3$ and $\epsilon^r_3$, and the inclination angle $i$.}
\label{fig2}
\end{figure*}

\begin{figure*}
\begin{center}
\includegraphics[type=pdf,ext=.pdf,read=.pdf,width=8.0cm]{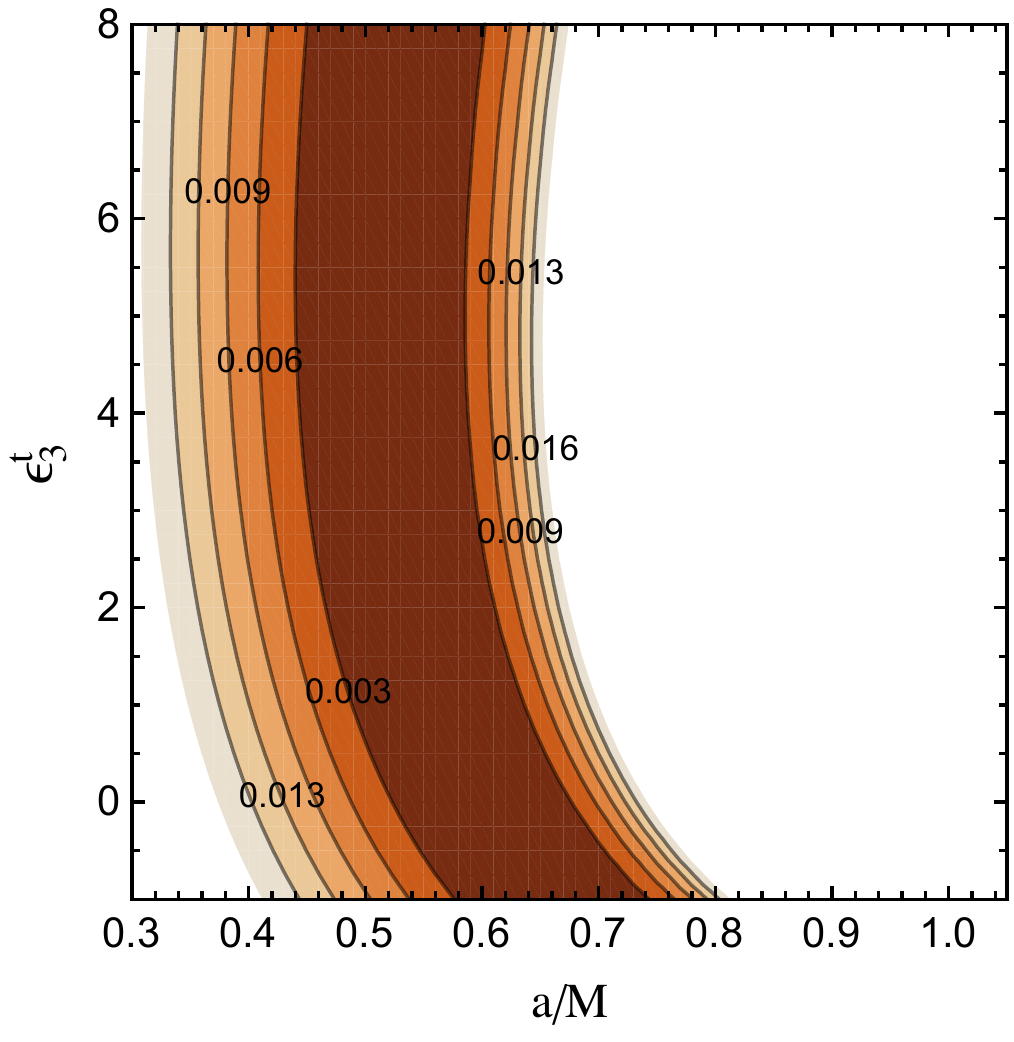}
\hspace{0.8cm}
\includegraphics[type=pdf,ext=.pdf,read=.pdf,width=8.0cm]{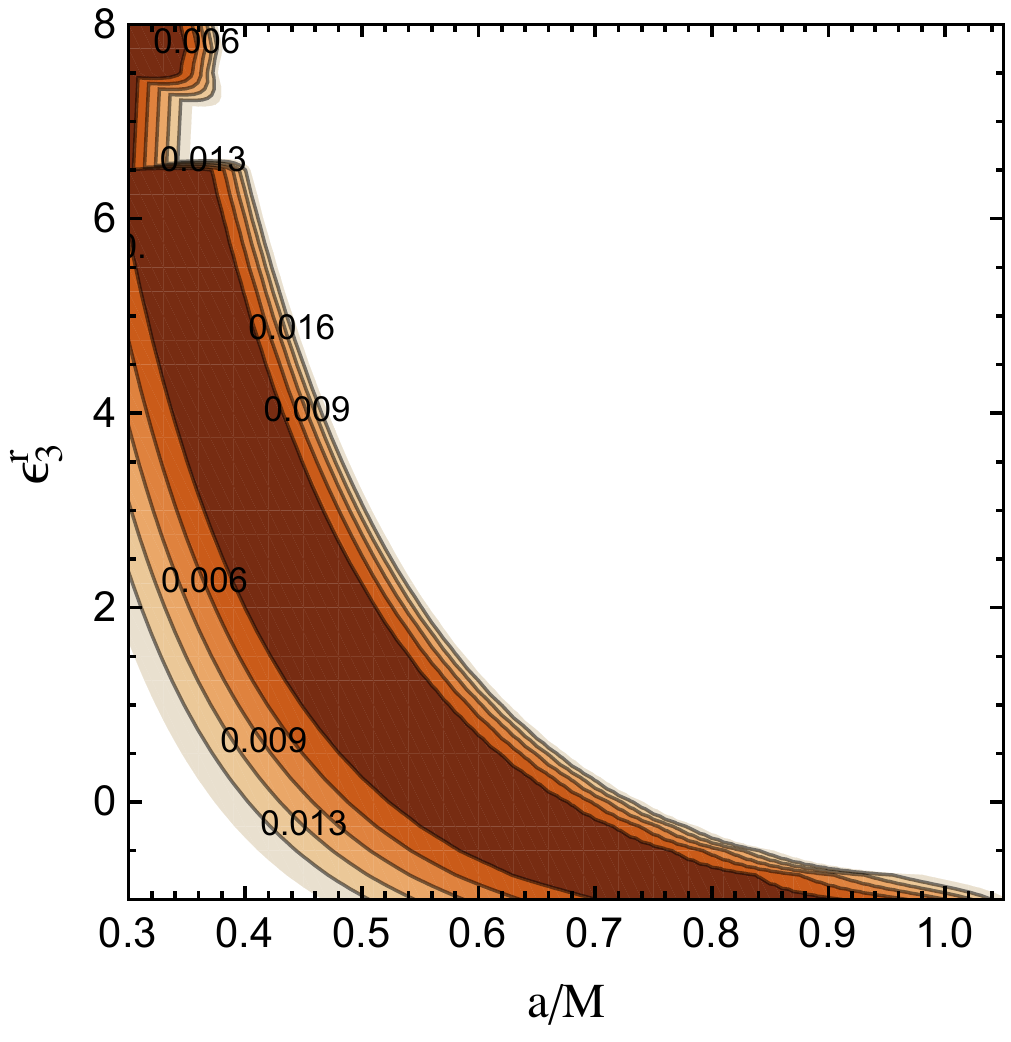} 
\end{center}
\caption{Contour maps of $S$. The reference model is a Kerr BH with the spin parameter $a_* = 0.6$ and observed with the inclination angle $i=80^\circ$.}
\label{fig3}
\vspace{0.8cm}
\begin{center}
\includegraphics[type=pdf,ext=.pdf,read=.pdf,width=8.0cm]{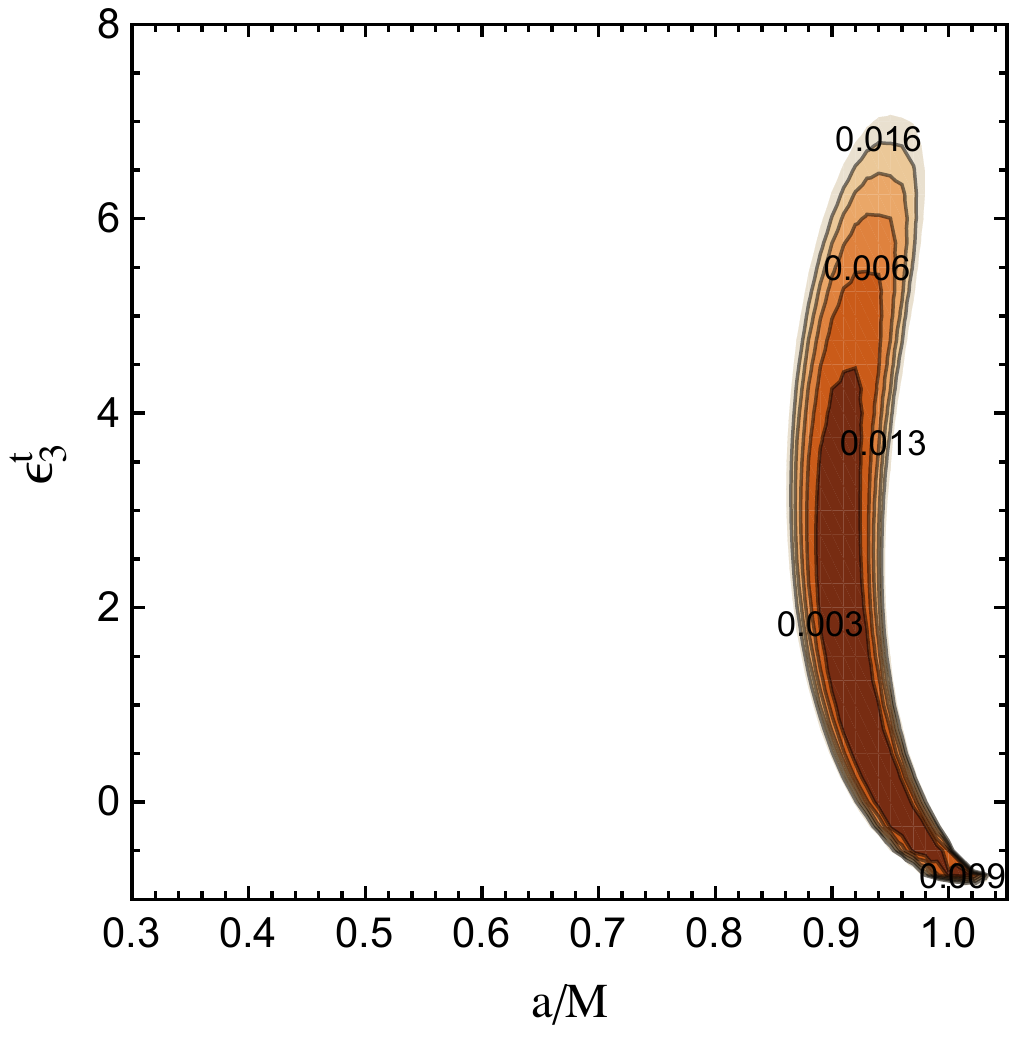}
\hspace{0.8cm}
\includegraphics[type=pdf,ext=.pdf,read=.pdf,width=8.0cm]{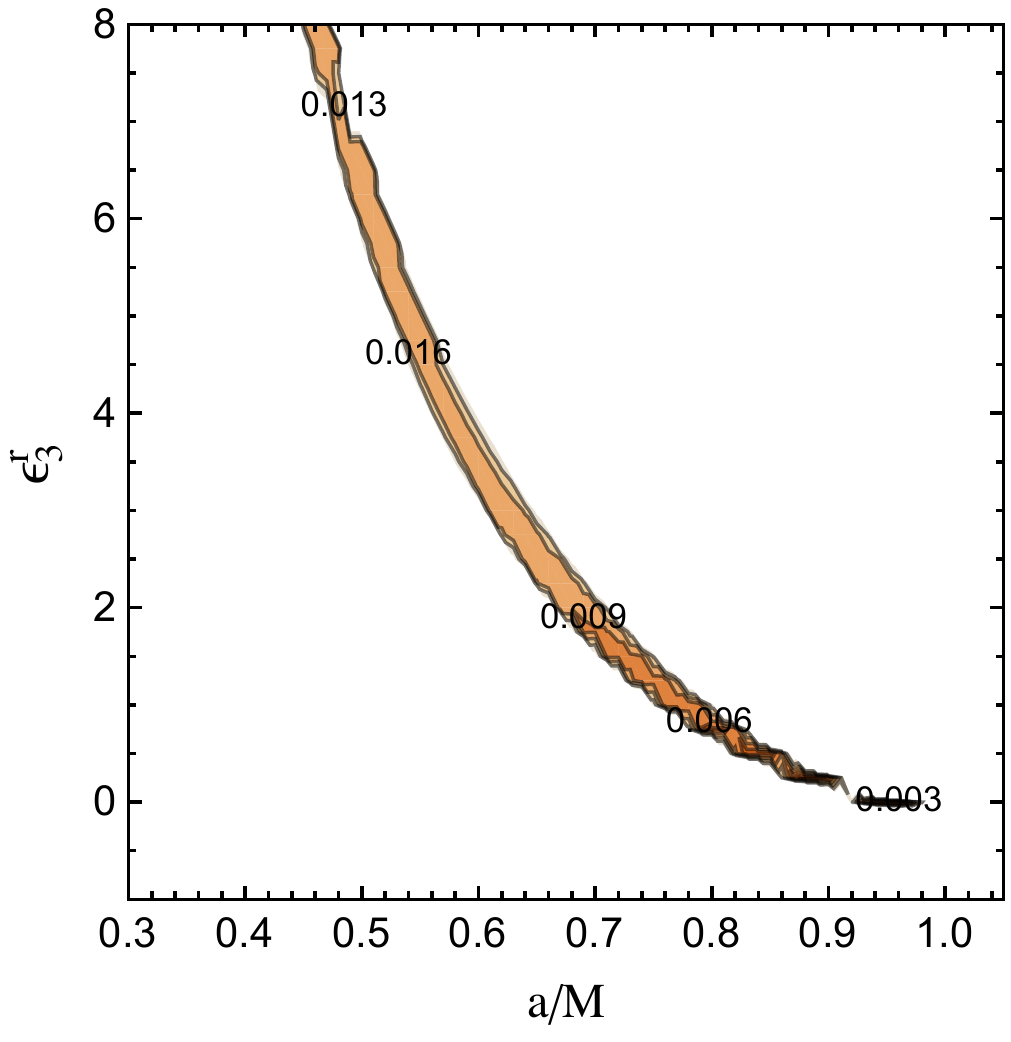} 
\end{center}
\caption{Contour maps of $S$. The reference model is a Kerr BH with the spin parameter $a_* = 0.95$ and observed with the inclination angle $i=80^\circ$.}
\label{fig4}
\end{figure*}

\begin{figure*}
\begin{center}
\includegraphics[type=pdf,ext=.pdf,read=.pdf,width=8.0cm]{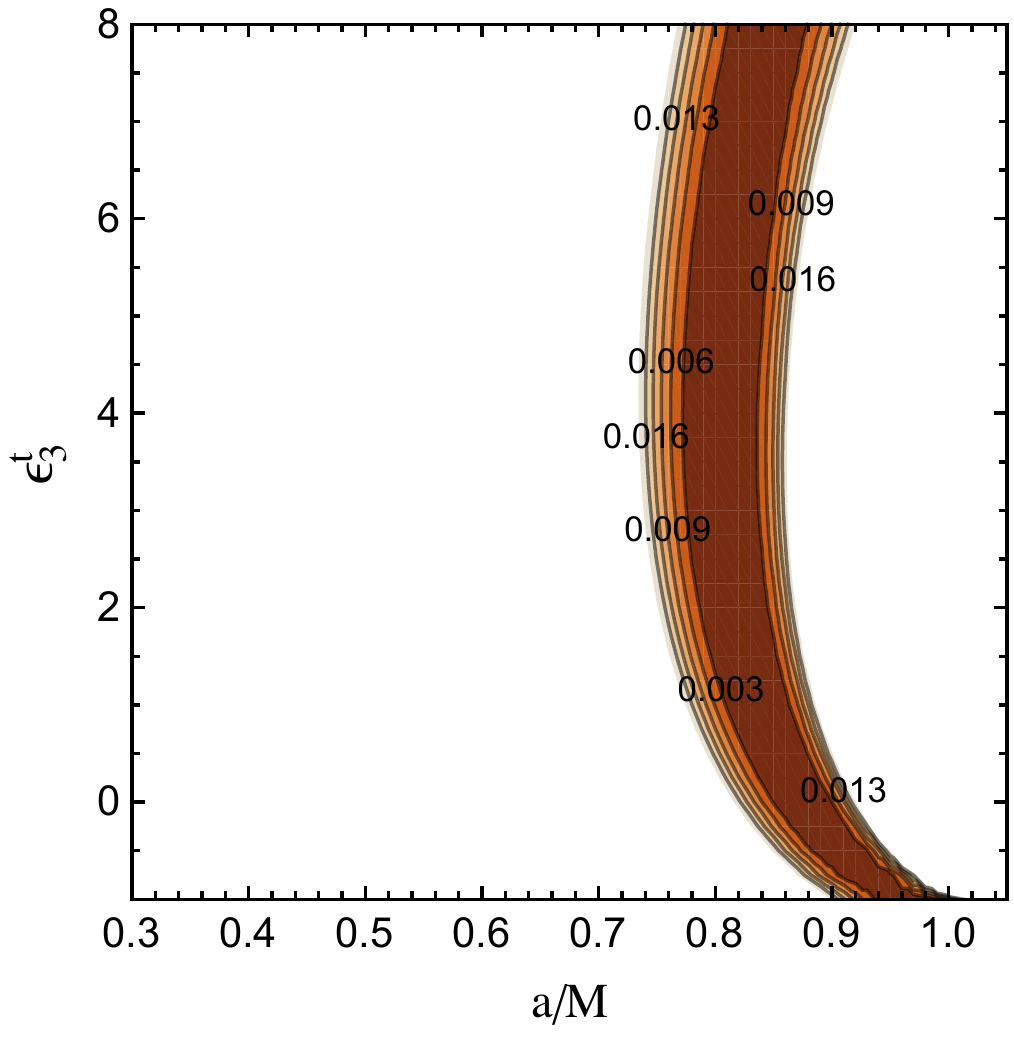}
\hspace{0.8cm}
\includegraphics[type=pdf,ext=.pdf,read=.pdf,width=8.0cm]{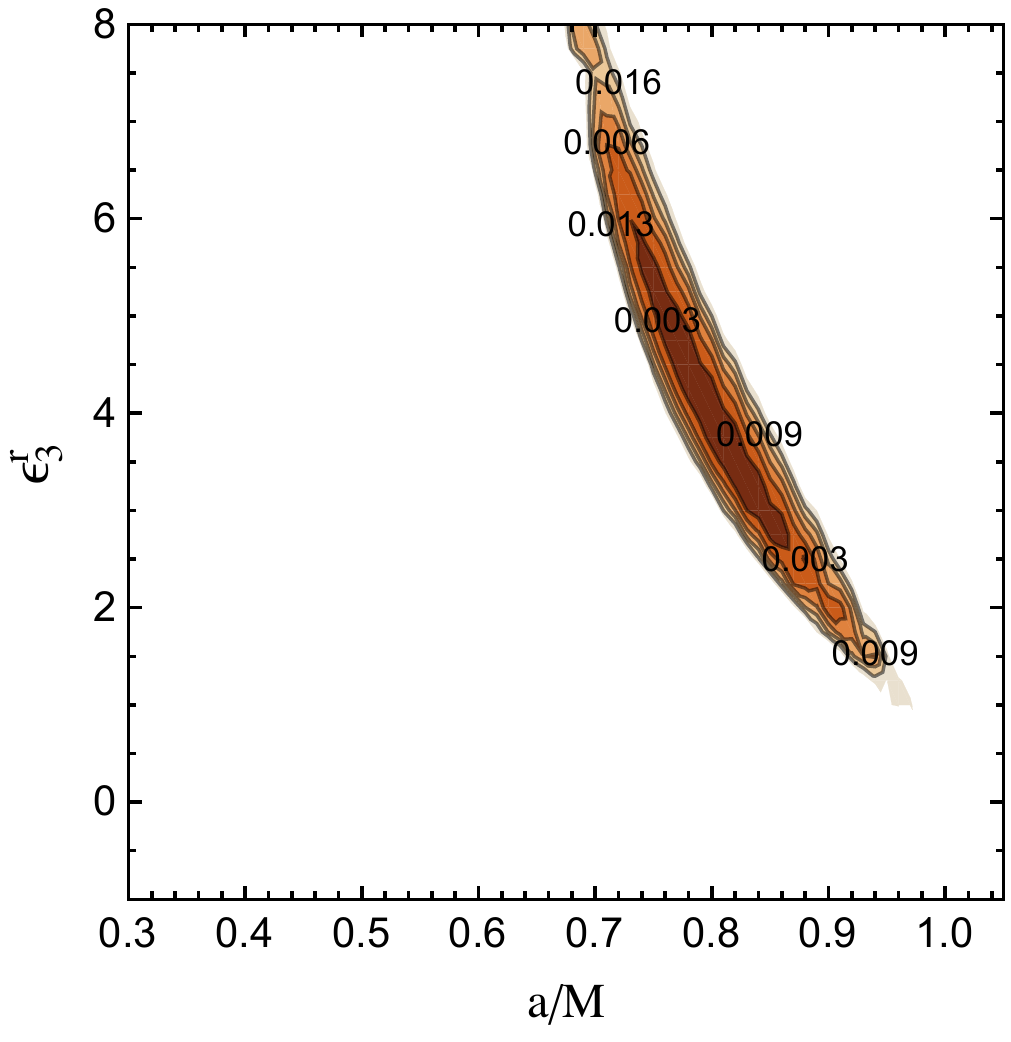} 
\end{center}
\caption{Contour maps of $S$. In the left panel, the reference model is a CPR BH with the spin parameter $a_* = 0.8$, the deformation parameters $\epsilon^t_3=4$ and $\epsilon^r_3=0$, and observed with the inclination angle $i=80^\circ$. In the right panel, the reference model is a CPR BH with the spin parameter $a_* = 0.8$, the deformation parameters $\epsilon^t_3=0$ and $\epsilon^r_3=4$, and observed with the inclination angle $i=80^\circ$.}
\label{fig5}
\end{figure*}

\section{Results \label{s-4}}

With the ingredients discussed in the previous section, we can compute the BH shadow on the image plane of the distant observer and the function $R(\phi)/R(0)$ for a specific set of the parameters $(a_*, \epsilon^t_3, \epsilon^r_3, i)$. Fig.~\ref{fig2} shows some examples. In the left panels, we consider the possibility of a non-vanishing $\epsilon^t_3$ and we assume $\epsilon^r_3 = 0$. In the right panels, we have the opposite case and $\epsilon^t_3 = 0$ while $\epsilon^r_3 \neq 0$.

The main impact of $\epsilon^t_3$ on the shadow is to alter its size: when $\epsilon^t_3 > 0$ the size of the shadow is smaller than that of a Kerr BH, when $\epsilon^t_3 < 0$ the size is larger. Such an effect can be understood by noting that $\epsilon^t_3$ represents a deformation of the metric coefficient $g_{tt}$, which regulates the intensity of the gravitational force. It is worth reminding that in the Newtonian limit $g_{tt} = -(1 + 2 \Phi)$, where $\Phi$ is the Newtonian gravitational potential, while all the other metric coefficients have the same form as in flat spacetime.

From the right panels in Fig.~\ref{fig2}, we see that a non-vanishing $\epsilon^r_3$ does not change the size of the shadow but affects its boundary, even if the effect is appreciable only on the side corresponding to corotating photon orbits (right sides in the shadows reported in Fig.~\ref{fig2}). In the case of high spin ($a_* = 0.9)$ and positive $\epsilon^r_3$ ($\epsilon^r_3 = 2$, 5), the boundary of the shadow present a peculiar shape. Such a feature is due to the fact that the photon capture sphere on that side is very close to the BH (a high spin and a positive $\epsilon^r_3$ make the gravitational force weaker and therefore the photons capture sphere is smaller) and the peculiar shape of the event horizon of these BHs. The latter is given by the larger root in
\be
\Delta + h^r a^2 \sin^2\theta = 0 \, ,
\ee
and it is thus only determined by $\epsilon^r_3$, not by $\epsilon^t_3$. As shown in Ref.~\cite{topo}, for high values of the BH spin the even horizon may have a non-trivial topology. The possible detection of a similar shadow would surely represent a clear indication of deviations from the Kerr metric.

To be more quantitative and figure out if and how different shadows can be distinguished, we proceed in the following way. We consider a ``reference model'', namely a BH with a specific set of spin, deformation parameters, and viewing angle. Given another BH with parameters $(a_*, \epsilon^t_3, \epsilon^r_3, i)$, we define the function 
\be
S(a_*, \epsilon^t_3, \epsilon^r_3, i) = \sum_k \left( 
\frac{R(a_*, \epsilon^t_3, \epsilon^r_3, i; \phi_k)}{R(a_*, \epsilon^t_3, \epsilon^r_3, i; 0)} - 
\frac{R^{\rm ref} (\phi_k)}{R^{\rm ref} (0)} \right)^2 \, . \nonumber\\
\ee
where $R(a_*, \epsilon^t_3, \epsilon^r_3, i; \phi_k)$ is the function $R$ at $\phi = \phi_k$, $\{ \phi_k \}$ is a set of angles $\phi$ for which we consider a measurement, and $R^{\rm ref} (\phi_k)$ is the function $R$ of the reference model. In what follows, we use 361 sample points, so $k$ runs from 0 to 360. The function $S$ is used to get a simple estimate of the similarity between two shadows. To have a rough idea of its meaning, we note that $\chi^2 \approx S/\sigma^2$, where $\sigma^2$ is the square of the error. Such a relation is only approximative, but it is enough for our purpose. For instance, if the shadow is determined with an uncertainty of 3\%, $\sigma \approx 0.03$, and $\chi^2 \approx 1000 \; S$. Since we do not introduce any noise in our treatment, $\Delta \chi^2 = \chi^2 - \chi^2_{\rm min} = \chi^2$. The contour levels $\Delta \chi^2 = 3.53$, 8.03, and 14.16, which correspond, respectively, to 68.3\%, 95.4\%, and 99.7\% confidence level for three degrees of freedom (the probability interval designated as 1-, 2-, and 3-standard deviations), become $S \approx 0.003$, 0.008, 0.014.

\begin{figure*}
\begin{center}
\includegraphics[type=pdf,ext=.pdf,read=.pdf,width=8.0cm]{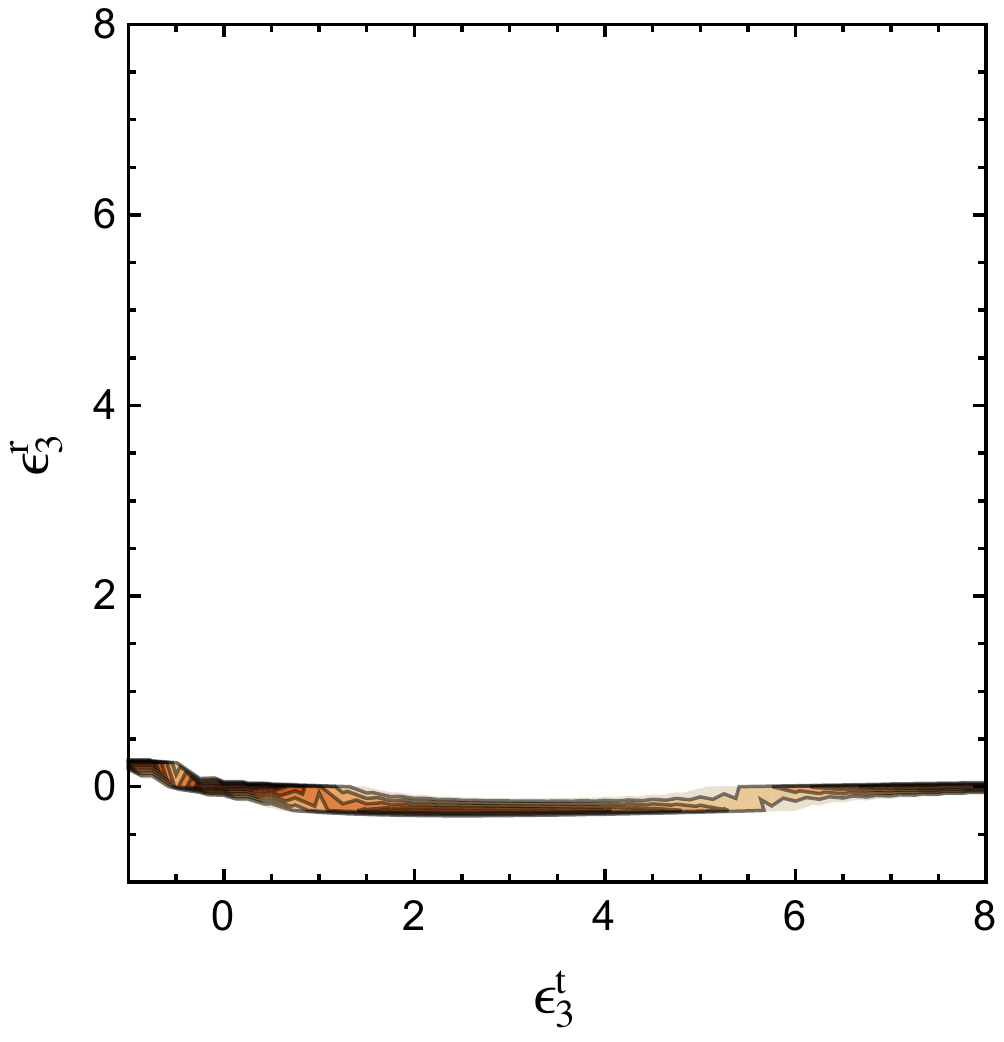}
\hspace{0.8cm}
\includegraphics[type=pdf,ext=.pdf,read=.pdf,width=8.0cm]{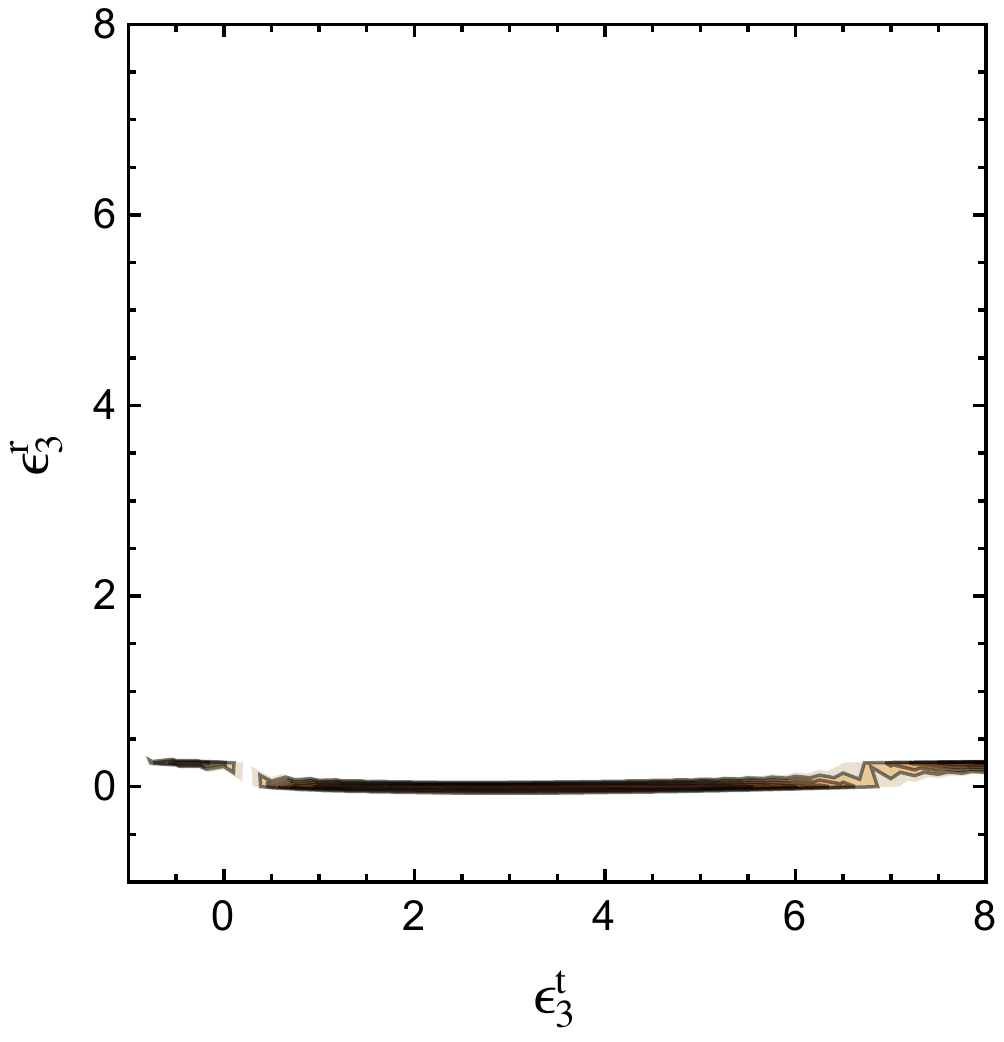} 
\end{center}
\caption{Contour maps of $S$. In the left panel, the reference model is a Kerr BH with the spin parameter $a_* = 0.9$ and observed with the inclination angle $i=80^\circ$. In the right panel, the reference model is a CPR BH with the spin parameter $a_* = 0.9$, the deformation parameters $\epsilon^t_3=4$ and $\epsilon^r_3=0$, and observed with the inclination angle $i=80^\circ$. Here we assume to know the value of the spin and we want to constrain $\epsilon^t_3$ and $\epsilon^r_3$. The contour levels are for $S=0.003$, 0.006, 0.009, 0.013, and 0.016 as in the other plots in this paper.}
\label{fig6}
\vspace{0.8cm}
\begin{center}
\includegraphics[type=pdf,ext=.pdf,read=.pdf,width=8.0cm]{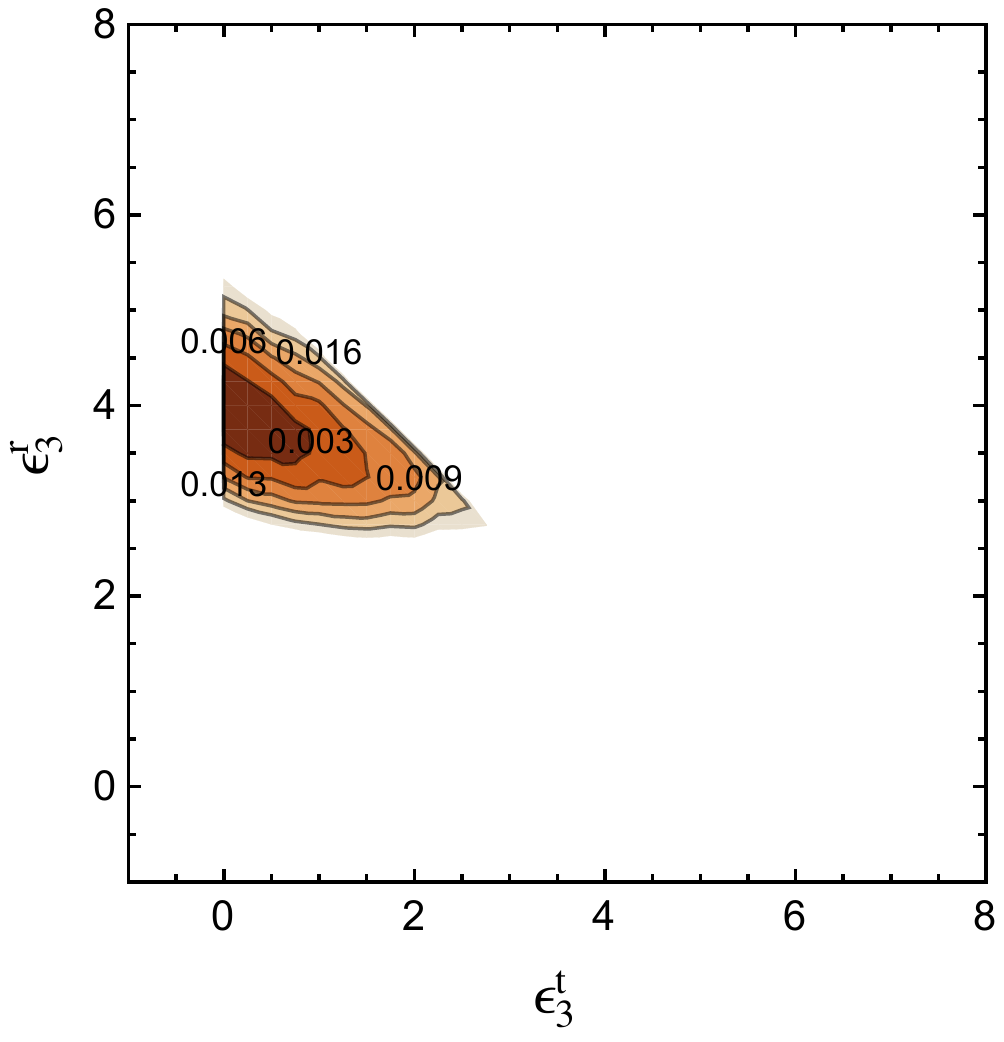}
\hspace{0.8cm}
\includegraphics[type=pdf,ext=.pdf,read=.pdf,width=8.0cm]{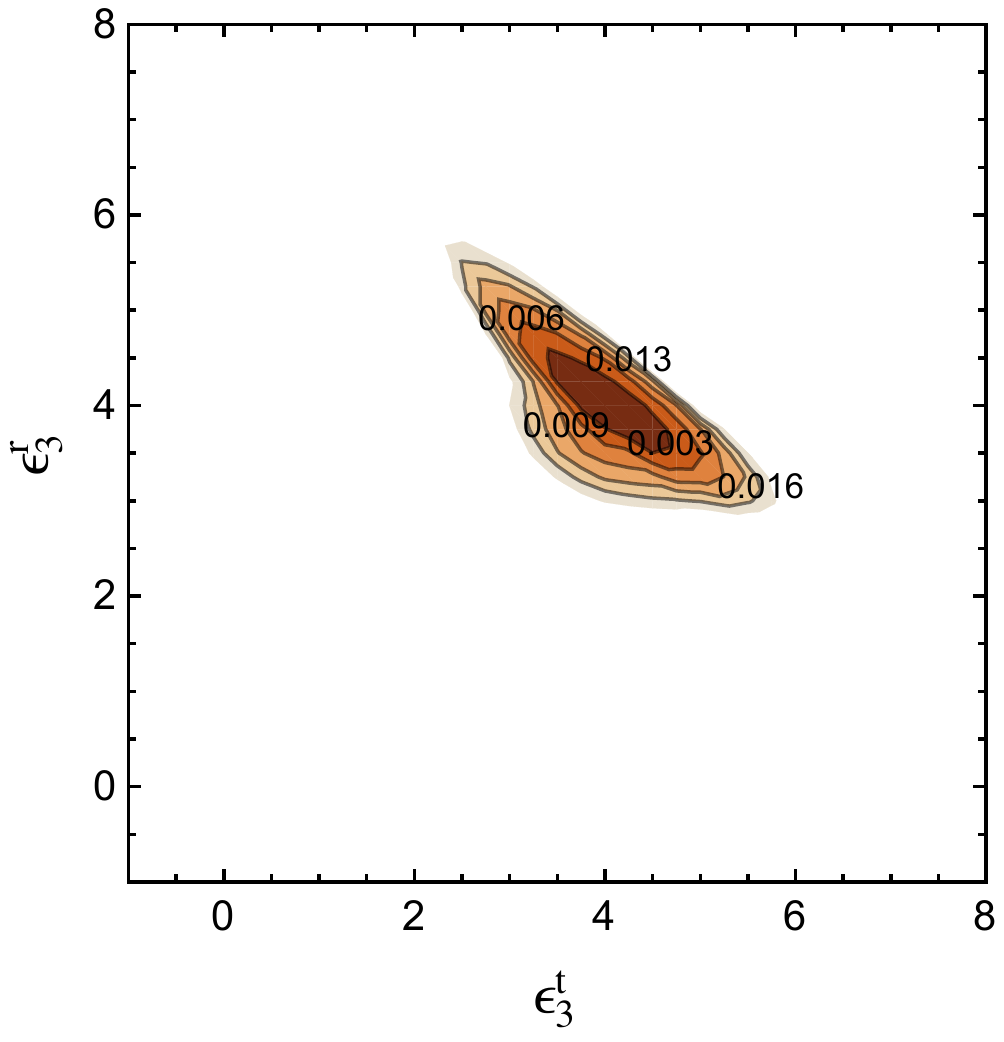} 
\end{center}
\caption{Contour maps of $S$. In the left panel, the reference model is a CPR BH with the spin parameter $a_* = 0.9$, the deformation parameters $\epsilon^t_3=0$ and $\epsilon^r_3=4$, and observed with the inclination angle $i=80^\circ$. In the right panel, the reference model is a CPR BH with the spin parameter $a_* = 0.9$, the deformation parameters $\epsilon^t_3=4$ and $\epsilon^r_3=4$, and observed with the inclination angle $i=80^\circ$. Here we assume to know the value of the spin and we want to constrain $\epsilon^t_3$ and $\epsilon^r_3$.}
\label{fig7}
\end{figure*}

Figs.~\ref{fig3}-\ref{fig7} show the contour levels of $S$ for different reference models. In Fig.~\ref{fig3}, the reference model has $a_* = 0.6$, $\epsilon^t_3 = \epsilon^r_3 = 0$ (Kerr BH), and $i = 80^\circ$. The left panel is to constrain $\epsilon^t_3$ assuming $\epsilon^r_3=0$, while the right panel is to constrain $\epsilon^r_3$ with the condition $\epsilon^t_3 = 0$. Here and in the other contour-plots of this paper, $i$ is a free parameter and we have selected the value that minimizes $S$. These plots clearly show that it is impossible to constrain the deformation parameters: we cannot exclude very large deviations from Kerr. We note that we are considering quite a large viewing angle of the reference model, and this should maximize the relativistic effects. This means that in the case of a lower viewing angle, it is even more difficult to constrain the metric.

Fig.~\ref{fig4} is devoted to the reference model $a_* = 0.95$, $\epsilon^t_3 = \epsilon^r_3 = 0$ (Kerr BH), and $i = 80^\circ$. As in the previous case, the left panel is for $\epsilon^t_3$, the right panel is for $\epsilon^r_3$. The main difference with Fig.~\ref{fig3} is that it is now possible to put a bound on $\epsilon^t_3$ and $\epsilon^r_3$, because the contour levels of $S$ are closed. We note, however, that this is an ideal case (large viewing angle, high spin parameter): we may not be so lucky and have SgrA$^*$ in this configuration.

Fig.~\ref{fig5} shows two cases in which the reference model is not a Kerr BH. In the left panel, we have $a_* = 0.8$, $\epsilon^t_3 = 4$, $\epsilon^r_3 = 0$, and $i = 80^\circ$. The conclusion is that we cannot constrain $\epsilon^t_3$ because the same shadow can be reproduced by a Kerr BH. The reference model in the right panels has $a_* = 0.8$, $\epsilon^t_3 = 0$, $\epsilon^r_3 = 4$, and $i = 80^\circ$. As we have seen in Fig.~\ref{fig2}, BH with positive $\epsilon^r_3$ can develop a shadow with a peculiar shape on the side of the corotaing photon orbits. This reference BH belongs to this class. No Kerr BH can mimic it and therefore the possible observation of a similar shadow could tell us that BH candidates are not the Kerr BHs of general relativity. We note, however, that such a value of the deformation parameter may be quite large if we think the metric in Eq.~(\ref{eq-m}) as a perturbation around the Kerr one, in which case $\epsilon^i_j$ should be much smaller than 1.

Lastly, we want to see if it is possible to constrain $\epsilon^t_3$ and $\epsilon^r_3$ at the same time and if there is a correlation between the measurements of these two parameters. For simplicity, we assume to know the spin parameter, for instance from the observations of pulsars~\cite{pulsar} or stars~\cite{stars}. In the left panel in Fig.~\ref{fig6}, the reference model has $a_* = 0.9$, $\epsilon^t_3 = \epsilon^r_3 = 0$ (Kerr BH), and $i = 80^\circ$. The fact we know the spin parameter allow a strong constraint on $\epsilon^r_3$, while the measurement of $\epsilon^t_3$ is more difficult. The conclusion is anyway that the correlation between the measurement of $\epsilon^t_3$ and $\epsilon^r_3$ is weak. In the right panel in Fig.~\ref{fig6}, the reference model has $a_* = 0.9$, $\epsilon^t_3 = 4$, $\epsilon^r_3 = 0$, and $i = 80^\circ$. Even in this case, the two measurements are only weakly correlated, $\epsilon^r_3$ can be well constrained, the estimate of $\epsilon^t_3$ is more difficult.

In the left panel in Fig.~\ref{fig7}, the reference model has $a_* = 0.9$, $\epsilon^t_3 = 0$, $\epsilon^r_3 = 4$, and $i = 80^\circ$\footnote{We note that the sharp cut at $\epsilon^t_3 = 0$ in the contour levels of $S$ is due to the fact that the metric is not defined beyond: $(1 + h^t)(1+h^r)$ becomes negative outside the event horizon.}. In the right panel in Fig.~\ref{fig7}, the reference model has $a_* = 0.9$, $\epsilon^t_3 = 4$, $\epsilon^r_3 = 4$, and $i = 80^\circ$. Since these BHs have relatively high spin and positive $\epsilon^r_3$, their shadow presents the feature discussed above associated to the strange shape of their even horizon and they can be distinguished from the Kerr BHs of general relativity. Contrary to the case with $\epsilon^r_3=0$, now it is easier to put a constraint on both $\epsilon^t_3$ and $\epsilon^r_3$ at the same time. The estimates of these two parameters is (anti)correlated.

\section{Concluding remarks \label{s-5}}

The apparent image of a BH surrounded by an optically thin emitting medium presents a shadow, which is a dark area over a brighter background. The boundary of the shadow corresponds to the apparent photon capture sphere and it is only determined by the spacetime geometry around the compact object and the viewing angle of the distant observer. An accurate detection of the shadow of a BH can thus provide information about the nature of the object and test the Kerr metric.

In this paper we have extended previous studies to use the BH shadow to test the Kerr metric. We have considered the CPR metric and calculated BH shadows for different values of the spin parameter $a_*$, the deformation parameters $\epsilon^t_3$ and $\epsilon^r_3$, and the viewing angle $i$. $\epsilon^t_3$ enters the metric coefficient $g_{tt}$ and therefore it can regulate the intensity of the gravitational force. The size of the BH shadow increases for $\epsilon^t_3 < 0$ and decreases for $\epsilon^t_3 > 0$. $\epsilon^r_3$ enters the metric coefficient $g_{rr}$ and therefore it determines the BH horizon. When $\epsilon^r_3 > 0$, fast-rotating BHs develop a topologically non-trivial horizon, which may imprint an observational signature in the BH shadow.

With the use of the $R$ function described in~\ref{sub}, we have compared the shadows of CPR BHs with different $a_*$, $\epsilon^t_3$, $\epsilon^r_3$, and $i$ to figure out if and how possible future measurements can test the Kerr metric. We have focused our attention to the optimistic case of a large viewing angle and set $i = 80^\circ$ for the reference model. Our results can be summarized as follows:
\begin{enumerate}
\item For a mid-rotating Kerr BH with the spin parameter $a_* = 0.6$, the measurement of its shadow can only provide an allowed region on the spin parameter -- deformation parameter plane. There is a fundamental degeneracy between the spin and possible deviations from the Kerr geometry and therefore it is possible to test the nature of the BH candidate only in the presence of independent measurements capable of breaking this degeneracy. The situation would be clearly worse with a lower value of the viewing angle.
\item For a fast-rotating Kerr BH with the spin parameter $a_* = 0.95$, it is potentially possible to constrain the deformation parameters. This is because high values of the spin and of the viewing angle maximize the relativistic effects around the BH. As the viewing angle decreases, it becomes more and more difficult to test the Kerr metric. In the case $i = 0^\circ$, all the shadows are just a circle, and therefore it is fundamentally impossible to get information on the spacetime metric. 
\item The shadow of some non-Kerr BHs presents some peculiar features. This may be the case of fast-rotating BHs with positive $\epsilon^r_3$: the even horizon of these objects is dramatically different from that of Kerr BHs and this leaves an observational signature on their shadow. The possible detection of a similar shadow would be enough to discover deviations from the Kerr geometry.
\item If we know the BH spin parameter from an independent measurement, we can try to constrain $\epsilon^t_3$ and $\epsilon^r_3$ at the same time. If the BH has high values of $a_*$ and $i$, and $\epsilon^r_3 = 0$, the detection of the shadow can well constrain $\epsilon^r_3$, while it is more difficult to measure $\epsilon^t_3$. The estimate of the two parameters is only weakly correlated. If $\epsilon^r_3 >0$, the peculiar shape of these shadows makes it easier a measurement of $\epsilon^t_3$ and $\epsilon^r_3$. In this case, the estimate of the two parameters is correlated.
\end{enumerate}


\begin{acknowledgments}
This work was supported by the NSFC grant No.~11305038, the Shanghai Municipal Education Commission grant for Innovative Programs No.~14ZZ001, the Thousand Young Talents Program, and Fudan University. M.G.-N. also acknowledges support from China Scholarship Council (CSC), grant No.~2014GXZY08.
\end{acknowledgments}


\end{document}